\documentclass[twocolumn,preprintnumbers,showpacs,amsmath,amssymb]{revtex4}
\usepackage{graphicx}
\usepackage{amssymb}

\begin{document}
\title{Comment on \textquotedblleft Complete insecurity of quantum protocols
for classical two-party computation\textquotedblright }
\author{Guang Ping He}
\email{hegp@mail.sysu.edu.cn}
\affiliation{School of Physics and Engineering, Sun Yat-sen University, Guangzhou 510275,
China}

\begin{abstract}
In a recent paper (Phys. Rev. Lett. 109, 160501 (2012). arXiv:1201.0849), it
is claimed that any quantum protocol for classical two-sided computation
between Alice and Bob can be proven completely insecure for Alice if it is
secure against Bob. Here we show that the proof is not sufficiently general,
because the security definition it based on is only a sufficient condition
but not a necessary condition.
\end{abstract}

\pacs{03.67.Dd, 03.67.Hk}
\maketitle

\newpage

Let us first look at the security definition in \cite{qbc61}. As stated in
the paragraph below its FIG. 1, let $\varepsilon \geq 0$ and
write $\rho \simeq _{\varepsilon }\sigma $ (i.e., $\rho $ is $\varepsilon $%
-close to $\sigma $) if the purified distance $\sqrt{1-(tr\sqrt{\sqrt{\rho }%
\sigma \sqrt{\rho }})^{2}}$\ between the density matrices $\rho $\ and $%
\sigma $\ is not greater than $\varepsilon $. Then a two-party
quantum protocol corresponding to a completely positive
trace-preserving (CPTP) map $\pi $ is defined as $\varepsilon
$-secure against dishonest Bob if for
any real adversary $B^{\prime }$ there exists an ideal adversary $\hat{B}%
^{\prime }$\ such that $[id_{R}\otimes \pi _{A,B^{\prime }}](\rho
_{UVR})\simeq _{\varepsilon }[id_{R}\otimes \mathcal{F}_{\hat{A},\hat{B}%
^{\prime }}](\rho _{UVR})$. Here $A$ denotes the real honest Alice, $%
B^{\prime }$ the dishonest Bob, and $\hat{A}$, $\hat{B}^{\prime }$ the ideal
versions. Both parties obtain an input (Alice's $u$ in register $U$ and
Bob's $v$ in register $V$) drawn from the distribution $p(u,v)$. $%
[id_{R}\otimes \pi _{A,B^{\prime }}](\rho _{UVR})$\ is the output state of
the protocol augmented by the reference $R$, where $\rho _{UVR}$\ is a
purification of $\sum\nolimits_{u,v}p(u,v)\left\vert u\right\rangle
\left\langle u\right\vert _{U}\left\vert v\right\rangle \left\langle
v\right\vert _{V}$. And $\mathcal{F}$\ is an ideal functionality which
measures the inputs and outputs orthogonal states that correspond to the
function values of the classical two-sided computation. Please see \cite%
{qbc61} for more detailed explanations of the notations.

In simple words, as can be seen from Sec. 1.6 of \cite{qi1087}
(i.e., Ref. [12] of \cite{qbc61}), the meaning of this definition
can be understood as follows. Let $\alpha $ and $\beta $ be the
physical systems accessible to
Alice and Bob, respectively. Denote the density matrices of $\alpha $, $%
\beta $ as $\rho _{\alpha }$, $\rho _{\beta }$\ when Bob plays honestly, or
as $\rho _{\alpha }^{\prime }$, $\rho _{\beta }^{\prime }$ when he applies a
certain cheating strategy. If there is $\rho _{\alpha }^{\prime }\simeq
_{\varepsilon }\rho _{\alpha }$,\ the cheating strategy will be
nearly undetectable to Alice so that Bob can pass the security checks
in the protocol successfully, while if there is $\rho _{\beta }^{\prime
}\simeq _{\varepsilon }\rho _{\beta }$,\ a dishonest Bob can hardly gain any
extra information other than what is accessible to an honest Bob. Then the
above security definition means that a protocol is secure against Bob if for
any cheating strategy, there is always $\rho _{\beta }^{\prime }\simeq
_{\varepsilon }\rho _{\beta }$. For simplicity, we call such a cheating
strategy as a type I strategy.

Obviously, if \textit{any} cheating strategy currently known or
potentially exists in the world belongs to type I, then the
corresponding protocol is surely secure. Thus it is a sufficient
condition for guaranteeing the security of a protocol. But it is
important to question whether the reversed statement is also true.
That is, if a protocol is secure, does it necessarily guarantee that
\textit{all} cheating strategies have to be type I strategies? In
fact, if there is a cheating strategy which does
not satisfy $\rho _{\alpha }^{\prime }\simeq _{\varepsilon }\rho _{\alpha }$%
, then it will be detectable to Alice, so that the protocol can
remain secure against Bob no matter $\rho _{\beta }^{\prime }\simeq
_{\varepsilon }\rho _{\beta }$\ is satisfied or not. We call
strategies satisfying neither $\rho _{\alpha }^{\prime }\simeq
_{\varepsilon }\rho _{\alpha }$ nor $\rho _{\beta }^{\prime }\simeq
_{\varepsilon }\rho _{\beta }$ as type II strategies. Actually, they
are no strangers to quantum cryptography. In many existing
protocols, there are security checks in which the parties agree to
continue with the protocols only when some conditions are met.
Otherwise they can choose to abort in the middle of the process, and
the protocols output \textquotedblleft fail\textquotedblright\
instead of the output obtained by honest players. This implies that
the protocols are designed against type II strategies. Thus it is
clear that the existence of type II strategies does not necessarily
hurt the security of protocols. If a protocol is secure, then both
types I and II strategies are possible. That is, \textquotedblleft
all cheating strategies belong to type I\textquotedblright\ is not
the necessary condition for a protocol to be secure. Therefore,
while the security definition in \cite{qbc61} is a true statement,
it cannot be used as \textquotedblleft a two-party quantum protocol
is $\varepsilon $-secure against Bob \textit{if\ and only if} for
any real adversary $B^{\prime }$ there exists an ideal adversary
$\hat{B}^{\prime }$\ such that $[id_{R}\otimes \pi
_{A,B^{\prime }}](\rho _{UVR})\simeq _{\varepsilon }[id_{R}\otimes \mathcal{F%
}_{\hat{A},\hat{B}^{\prime }}](\rho _{UVR})$\textquotedblright,
since the reversed statement \textquotedblleft for any real
adversary $B^{\prime }$, there exists an ideal adversary
$\hat{B}^{\prime }$\ such that $[id_{R}\otimes \pi
_{A,B^{\prime }}](\rho _{UVR})\simeq _{\varepsilon }[id_{R}\otimes \mathcal{F%
}_{\hat{A},\hat{B}^{\prime }}](\rho _{UVR})$ if the protocol is
$\varepsilon $-secure against Bob\textquotedblright\ is not true.
There can be type II strategies which are not $\varepsilon $-close
to any ideal adversary.

Now back to the no-go proof for two-sided computation in
\cite{qbc61}. In brief, the key starting points of the proof are as
follows. Suppose that there is a quantum protocol for classical
two-sided computation which is already assumed to be secure against
a dishonest Bob. To prove that it must be insecure against Alice, in
the paragraph before Eq. (1) of \cite{qbc61}, the following cheating
strategy of Bob is considered. He plays the honest but purified
strategy and outputs the purification of the protocol (register
$Y_{1}^{\prime }$) and the output values $f(u,v)$\ (register $Y$).
We call it strategy $B_{0}^{\prime }$ hereafter. Since the protocol is $\varepsilon $%
-secure against Bob, in the opinion of \cite{qbc61} there exists a
secure state $\sigma _{RX\tilde{V}Y^{\prime }}$ satisfying $\sigma
_{RXY^{\prime }}\simeq _{\varepsilon }\rho _{RXY^{\prime }}$, where
$Y^{\prime }=Y_{1}^{\prime }Y$. Applying Uhlmann's theorem on
$\sigma _{RXY^{\prime }}\simeq _{\varepsilon }\rho _{RXY^{\prime
}}$, Eq. (1) of \cite{qbc61} can be obtained, which further leads to
the rest part of the no-go proof.

However, according to our above discussion on the security
definition, \textquotedblleft the protocol is $\varepsilon $-secure
against Bob\textquotedblright\ does not necessarily guarantees that
\textquotedblleft all cheating strategies (including strategy
$B_{0}^{\prime }$) must be type I strategies\textquotedblright ,
because the latter statement is not the necessary condition of the
former. If $B_{0}^{\prime }$ belongs to type II, then the protocol
can still be secure against Bob, while the equation $\sigma
_{RXY^{\prime }}\simeq _{\varepsilon }\rho _{RXY^{\prime }}$ no
longer holds. Consequently, Eq. (1) does not necessarily remain
valid so that the no-go proof will lose its base. Thus we can see
that the proof in \cite{qbc61} may apply to a protocol for which
$B_{0}^{\prime }$ can be proven to be a type I strategy (given that
all other features of the protocols studied in \cite{qbc61} are also
met). But it is not sufficient general to cover all protocols, since
there is no evidence (at least not provided in \cite{qbc61}) showing
that $B_{0}^{\prime }$ always\ has to be a type I strategy for any
protocol potentially exists. By designing proper security checks
which can make $B_{0}^{\prime }$ appear as a type II strategy, it is
possible to build protocols not covered by the proof in
\cite{qbc61}. Therefore, the door for finding secure quantum
protocols for classical two-party computation is not closed
completely.

The work was supported in part by the NSF of China under grant No. 10975198,
the NSF of Guangdong province, and the Foundation of Zhongshan University
Advanced Research Center.

\end{document}